\documentclass[twocolumn]{aastex631}

\usepackage{amsmath,amssymb,amsxtra,amsfonts}   
\usepackage{txfonts}
\usepackage{xspace}

\newcommand{\CII}{[C\,\textsc{ii}]}
\newcommand{\Msun}{$M_\odot$\xspace}
\newcommand{\Mstar}{$M_\ast$\xspace}

\newcommand{\targetname}{A2744-DSG-z3}

\begin{document}

\title{The Identification of a Dusty Multiarm Spiral Galaxy at $z=3.06$ with JWST and ALMA}

\author[0000-0003-0111-8249]{Yunjing Wu}
\affiliation{Department of Astronomy, Tsinghua University, Beijing 100084, China\email{zcai@mail.tsinghua.edu.cn}}
\affiliation{Steward Observatory, University of Arizona, 933 N Cherry Ave, Tucson, AZ 85721, USA}

\author[0000-0001-8467-6478]{Zheng Cai}
\affiliation{Department of Astronomy, Tsinghua University, Beijing 100084, China\email{zcai@mail.tsinghua.edu.cn}}

\author[0000-0002-4622-6617]{Fengwu Sun}
\affiliation{Steward Observatory, University of Arizona, 933 N Cherry Ave, Tucson, AZ 85721, USA}

\author[0000-0002-1620-0897]{Fuyan Bian}
\affiliation{European Southern Observatory, Alonso de C\'{o}rdova 3107, Casilla 19001, Vitacura, Santiago 19, Chile}

\author[0000-0001-6052-4234]{Xiaojing Lin}
\affiliation{Department of Astronomy, Tsinghua University, Beijing 100084, China\email{zcai@mail.tsinghua.edu.cn}}

\author[0000-0001-5951-459X]{Zihao Li}
\affiliation{Department of Astronomy, Tsinghua University, Beijing 100084, China\email{zcai@mail.tsinghua.edu.cn}}

\author[0000-0001-6251-649X]{Mingyu Li}
\affiliation{Department of Astronomy, Tsinghua University, Beijing 100084, China\email{zcai@mail.tsinghua.edu.cn}}

\author[0000-0002-8686-8737]{Franz E. Bauer}
\affiliation{Instituto de Astrofısica, Facultad de Fısica, Pontificia Universidad Catolica de Chile Av. Vicuna Mackenna 4860, 782-0436
Macul, Santiago, Chile}
\affiliation{Millennium Institute of Astrophysics, Nuncio Monse{\~n}or S\'otero Sanz 100, Providencia, Santiago, Chile}

\author[0000-0003-1344-9475]{Eiichi Egami}
\affiliation{Steward Observatory, University of Arizona, 933 N Cherry Ave, Tucson, AZ 85721, USA}

\author[0000-0003-3310-0131]{Xiaohui Fan}
\affiliation{Steward Observatory, University of Arizona, 933 N Cherry Ave, Tucson, AZ 85721, USA}

\author[0000-0003-3926-1411]{Jorge Gonz\'alez-L\'opez}
\affil{Las Campanas Observatory, Carnegie Institution of Washington, Casilla 601, La Serena, Chile}
\affil{N\'ucleo de Astronom\'ia de la Facultad de Ingenier\'ia y Ciencias, Universidad Diego Portales, Av. Ej\'ercito Libertador 441, Santiago, Chile}

\author[0000-0002-1815-4839]{Jianan Li}
\affiliation{Department of Astronomy, Tsinghua University, Beijing 100084, China\email{zcai@mail.tsinghua.edu.cn}}

\author[0000-0002-7633-431X]{Feige Wang}
\altaffiliation{NHFP Hubble Fellow}
\affiliation{Steward Observatory, University of Arizona, 933 N Cherry Ave, Tucson, AZ 85721, USA}

\author[0000-0001-5287-4242]{Jinyi Yang}
\altaffiliation{Strittmatter Fellow}
\affiliation{Steward Observatory, University of Arizona, 933 N Cherry Ave, Tucson, AZ 85721, USA}

\author[0000-0002-0427-9577]{Shiwu Zhang}
\affiliation{Department of Astronomy, Tsinghua University, Beijing 100084, China\email{zcai@mail.tsinghua.edu.cn}}

\author[0000-0002-3983-6484]{Siwei Zou}
\affiliation{Department of Astronomy, Tsinghua University, Beijing 100084, China\email{zcai@mail.tsinghua.edu.cn}}

\begin{abstract}
Spiral arms serve crucial purposes in star formation and galaxy evolution. In this paper, we report the identification of ``A2744-DSG-$z3$”, a dusty, multiarm spiral galaxy at $z=3.059$ using the James Webb Space Telescope (JWST) NIRISS imaging and grism spectroscopy. \targetname\ was discovered as a gravitationally lensed sub-millimeter galaxy with ALMA. This is the most distant stellar spiral structure seen thus far, consistent with cosmological simulations which suggest $z\approx3$ as the epoch when spirals emerge. Thanks to the gravitational lensing and excellent spatial resolution of JWST, the spiral arms are resolved with a spatial resolution of $\approx290$\,pc. Based on SED fitting, 
the spiral galaxy has a de-lensed star formation rate of $85\pm30 \ M_{\odot}$ yr$^{-1}$, and a stellar mass of $\approx10^{10.6}\ M_{\odot}$, indicating that A2744-DSG-$z3$ is a main-sequence galaxy. After fitting the spiral arms, we find a stellar effective radius ($R_{e, \rm{star}}$) {of} $5.0\pm1.5$ kpc. 
{Combing with ALMA measurements,} we find that the effective radii ratio between dust and stars is $\approx0.4$, similar to {those} of massive SFGs at $z\sim2$, indicating a compact dusty core in \targetname. Moreover, this galaxy appears to be living in a group environment: including \targetname, at least three galaxies at $z=3.05 - 3.06$ {are} spectroscopically confirmed by JWST/NIRISS and ALMA, residing within a lensing-corrected projected scale of $\approx 70$ kpc. This, along with the asymmetric brightness profile, further suggests that the spiral arms may be triggered by minor merger events at $z\gtrsim3$. 

\end{abstract}



\section{Introduction} \label{sec:intro}

Spiral structures are common among {galaxies in the local Universe}. Some exhibit grand design {morphology}, in which prominent and well-defined spiral arms can be determined, while other galaxies show multi-arm or flocculent spirals, with subtler structural features \citep{Elmegreen2011, Elmegreen2014}.
Extensive efforts have been conducted to investigate the morphology, to understand the spiral/disk formation mechanism, and to study the relation between the spiral and star formation \citep[e.g.,][]{Conselice2014, Lin1964}. 
Nevertheless, when and how large spirals emerged in the early Universe is largely unknown. 
 
\begin{figure*}
    \centering
	\includegraphics[width=0.9\textwidth]{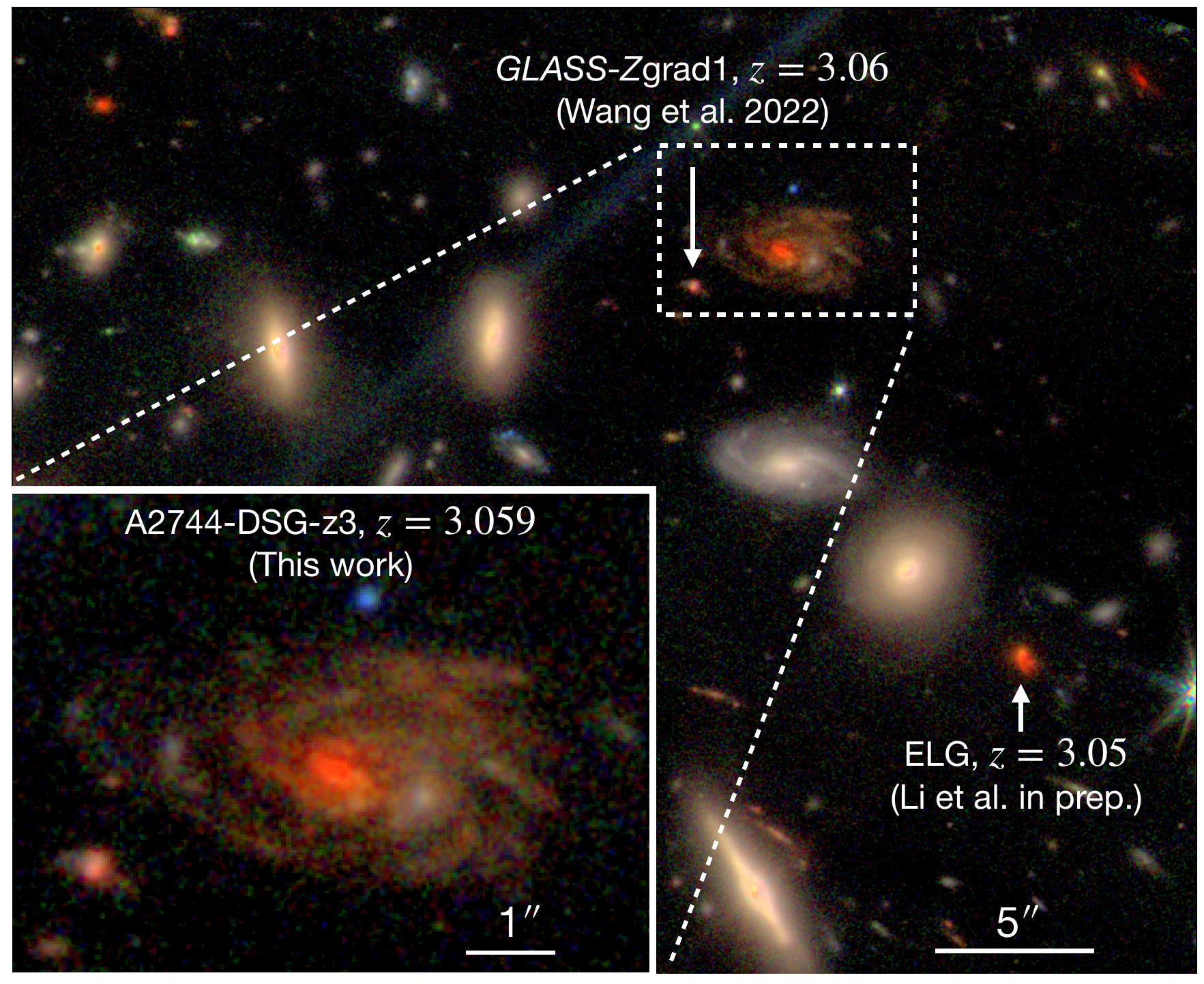}
	\caption{Dusty spiral galaxy (DSG) at $z=3.059$. The composite RGB map of the target \targetname\ is based on 3-band JWST/NIRISS imaging observations with pixel scale of 0.03\arcsec\ (blue: F115W, green: F150W, red: F200W). 
	The target \targetname\ was initially reported by the ALMA Frontier Fields survey \citep[][A2744-ID03 in their papers]{GL17,Laporte17} as a SMG (see also \citealt{Sun2022}).
	Meanwhile, the target $GLASS$-${\rm Z}$grad1, reported in \citet{Wang2022}, is a star-forming galaxy with stellar mass (\Mstar) of $\sim10^9$\Msun\ at {the} same redshift. The impact parameter between these two galaxies {is} $\approx3.1\arcsec$. {Another} emission-line galaxy is confirmed by NIRISS grism and ALMA as well at $z=3.05$ (Li et al. in prep.) with the projected separation of 15\arcsec.
	This galaxy was also reported as a SMG with $M_{\rm dust}\simeq10^{8.6}$\Msun\  \citep{Laporte17,Sun2022}. } 
	\label{spiral_arm_rgb}
\end{figure*}

Several studies have been conducted to search for spiral structures at $z\gtrsim2$, the peak of cosmic star formation, and to study the initial formation and evolution of the spiral morphology \citep[e.g.,][]{Margalef_Bentabol2022, Chen2022, Fudamoto2022}. 
With ALMA, \citet{Tsukui2021} discovered a cold, \CII-emitting rotating gaseous disk that can be interpreted as either spiral arms or tidal structures. Nevertheless, the stellar spiral structures are even rarer at high-$z$ Universe. Only a small number of stellar spiral structures 
are identified at $z > 2$ that have been spectroscopically confirmed {including} 
HDFX 28 at $z = 2.011$ \citep{Dawson2003}, Q2343-BX442 at $z = 2.18$ \citep{Law2012}, 
and RS14, which is recently shown by JWST observations at $z=2.46$ \citep{Fudamoto2022}. 
No stellar spiral structure {has yet been} confirmed at $z>3$. 

Currently, it is still unclear whether 
the rare appearance of spirals at high-$z$ is mainly 
due to the spiral formation or surface brightness dimming. 
At $z>2$, galaxies are observed to be clumpier and more irregular \citep[e.g.,][]{Elmegreen2009}.
\citet{Law2009} suggest that the irregularity at $z\gtrsim2$ can be explained, because 
the disks could be dynamically hot, and hot disks could yield more clumps  
than spiral structures \citep{Conselice2005}. 
Simulations further suggest that clumpy galaxies may transition into spirals. 
Nevertheless, at high-$z$, direct observational constraints linking gaseous clumps and spirals are lacking, while our understanding of how spirals may impact star formation remains unclear. 
For example,  
whether spirals at $z>2$ can enhance the star formation or not are still under debated \citep[e.g.,][]{Elmegreen2002, Moore2012}.
All these questions or debates can be addressed by identification of more spirals across 
cosmic time and studying their star formation properties. 
Such observations can also answer the long-standing question of 
when and where spiral structures start to emerge 
in the early Universe, as well as what their evolutionary endpoints at present time might be. 

In this paper, we present our identification of a stellar spiral at $z>3$. 
This galaxy is reported as a submillimeter galaxy (SMG) with the ALMA Frontier Field survey \citep{GL17,Laporte17}, and recovered recently with the ALMA Lensing Cluster Survey \citep[ALCS;][]{Sun2022} with updated spectroscopic redshift $z=3.06$. 
Nevertheless, ALMA spatial resolution ($\approx 0\farcs6-$1\farcs0) is  insufficient to resolve most spiral-like structures for galaxies at $z\gtrsim3$. 
JWST has a near-infrared (NIR) capability that allows us to probe the stellar properties of high-$z$ galaxies. 
{The high spatial resolution ($\approx 0.06\arcsec$) enables us to fully resolve the spiral structures.}
{We named this galaxy as \targetname\ in the following of the paper, where DSG is the acronym of ``dusty spiral galaxy".}

This paper is organized as follows: {Section~\ref{sec:data} describes} details of data and reduction {technique}. Section~\ref{sec:analysis} presents the spectra of \targetname\ obtained from ALMA and JWST/NIRISS and a stellar-mass surface density map. {In Section~\ref{sec:discussion}, we provide further discussion} of our observations. In this paper, we assume a flat cosmological model with $\rm{\Omega_{M} = 0.3, \Omega_{\Lambda} = 0.7}$ and $\rm{H_0 = 70\ km\ s^{-1}\ Mpc^{-1}}$, $1\arcsec=7.7$ kpc at $z=3.06$. 
The Glafic lensing model is adopted \citep{Oguri2010}.

\begin{figure*}
    \centering
	\includegraphics[width=0.95\textwidth]{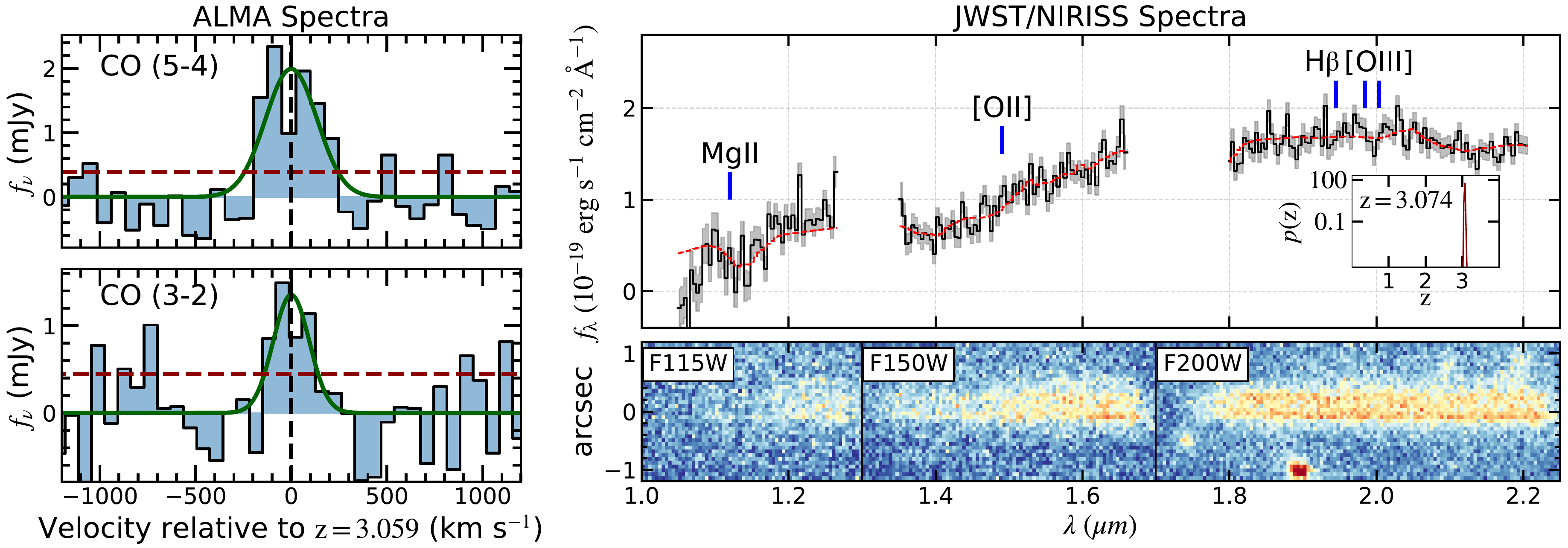}
	\caption{Determinations of the redshift of DSG-$z3$. {\bf Left}: ALMA CO (5-4) and (3-2) spectra of \targetname. The dashed red line shows 1-$\sigma$ noise spectra, while the dashed black line represents the systemic velocity. Dark-green curves show the best-fit Gaussian models of the spectra, indicating the CO-based redshift of $z_{\rm CO}=3.059\pm0.001$. {\bf Right}: Top panel shows the optimally extracted 1D spectra from NIRISS 3-band grism observations. Black histogram denotes the observed flux density, with gray shadows showing $1-\sigma$ errors. The dashed red lines are the \textsc{Grizli} modelled spectra containing continuum and nebular emission templates obtained by best SED fitting. The vertical blue lines mark the location of common emission lines ($[\textrm{O}~\textsc{iii}]\xspace,\textrm{H}\beta,[\textrm{O}~\textsc{ii}]\xspace$ and $\textrm{Mg}~\textsc{ii}\xspace$) that are not detected in DSG-$z3$. The subplot at bottom right shows the probability distribution function (PDF) of redshifts from SED-fitting, indicating that the best grism redshift of this target is $z\approx3.07$. Bottom panels are the joint two-dimensional spectra covered by NIRISS three filters (F115W, F150W and F200W).
    }
	\label{Spec}
\end{figure*}

\section{DATA} \label{sec:data}

\subsection{JWST/NIRISS imaging \& grism observations}
The NIRISS data were obtained from the GLASS {\it JWST} Early Release Science Program \citep{Treu2022} with direct imaging in F115W, F150W, and F200W and grism spectroscopic observations. 
The direct-imaging exposure time is $2834.504\,\mathrm{s} \simeq 0.79\,\mathrm{hours}$ per each band.
The data were reduced using the standard JWST pipeline\footnote{\href{https://github.com/spacetelescope/jwst}{https://github.com/spacetelescope/jwst}} v1.6.2 with calibration reference files ``\texttt{jwst\_0944.pmap}''.
1/f noise (see \citealt{Schlawin20}) was modeled and removed using the code \texttt{tshirt}\footnote{\href{https://github.com/eas342/tshirt}{https://github.com/eas342/tshirt}}, and the ``snowball'' artifacts from cosmic rays \citep{Rigby22} were identified and masked.
The world coordinate system of mosaicked images were registered using the Pan-STARRS1 catalog \citep{panstarrsdr1}. 
The pixel scale of final mosaicked images is resampled to 0.03\arcsec\ with \texttt{pixfrac}\,$=$\,0.8.

We reduced the grism data using \textsc{Grizli} \footnote{\url{https://github.com/gbrammer/grizli/}} \citep[Version 1.6,][]{Brammer_grizli_2022}. The 2D grism spectra were drizzled using a pixel scale 0.065\arcsec. The contamination was subtracted by forward modeling of the full field-of-view (FoV) grism images. We {derived} reliable spectroscopic redshifts via spectral template fitting based on a library of spectral energy distributions (SED). 

\subsection{ALMA CO observations \& the determination of redshift}
Band 3 and 4 line-scan observations of our target are available in the ALMA archive (Program 2017.1.01219.S; PI: Bauer).
The frequency ranges from 84.10--99.97\,GHz and 139.51--155.38\,GHz,
covering CO (3-2) and (5-4)  emission lines (rest-frame: 345.7960/576.2679 GHz) at $z=3.06$ as reported in \citet{Sun2022}.
The on-source time of these observations is 12.5 and 10.0 $\mathrm{mins}$.
The spectra were directly extracted by rebinning the channel width to 74/69 km/s for CO(3-2) and CO(5-4) of available data products, using the CASA package v6.3.0 \citep{McMullin2007}.

\begin{table}
\centering
\caption{Physical properties of \targetname. }
\begin{tabular}{lc}
\hline
Galaxy & \targetname \\
\hline
R.A. (deg) & 3.58502 \\
Dec  (deg) & --30.38181 \\
$z_{\rm spec}$     & 3.059 \\
$\mu$   & 2.45  \\
\hline
\multicolumn{2}{c}{Physical parameters}   \\
Inclination (deg)            &   28  \\
Pitch angle (deg)            &   $34\pm13$  \\
Hubble type            &    Sc  \\
${\rm SFR}$ ($M_\sun\ {\rm yr^{-1}}$)$^{\rm a}$ & $85\pm30$ \\
$\log [M_{\rm star}$/($M_\sun$)] & $10.55$ \\
$\log [M_{\rm dust}$ /($\mu^{-1}M_{\odot}$)]$^{\rm a}$ & $8.70 \pm 0.10$ \\
$R_{\rm e,\ star}$ (kpc) & $5.0\pm1.5$\\
$R_{\rm e,\ dust}$ (kpc)$^{\rm b}$ & 2.04 $\pm$ 0.26\\
\hline
\end{tabular}
\tablecomments{a. Reference of SFR and dust mass is from \citet{Sun2022}, 
b. De-lensed result. Reference of effective radius of dust continuum: Sun et al. (in prep.).}
\label{tab:info}
\end{table}

\section{Analysis \& Results} \label{sec:analysis}
In this letter, we focus on the physical properties of \targetname. 
Detailed information is listed in Table~\ref{tab:info}.
Figure~\ref{spiral_arm_rgb} shows the color composite map of this target. 
We note that there are two companion galaxies in the same field 
which could be related to \targetname. 
The first one is a star-forming dwarf galaxy close to \targetname, denoted as $GLASS$-${\rm Z}$grad1 \citep{Wang2022}.  
The {angular} offset is $\approx 3.1\arcsec$ on the image plane and then is de-lensed to $\approx 2\arcsec$.
The second companion galaxy is on the southwest side of \targetname. The observed angular separation is $\approx15\arcsec$, corresponding to a lensing-reconstructed impact parameter of $\approx 9.6\arcsec$ (i.e., $\approx74$ kpc at $z=3$; Li et al. in prep). 
The redshift of \targetname\ is determined by ALMA CO observations (see the left panel of Figure~\ref{Spec}). The redshifts of two companion galaxies are confirmed by the JWST/NIRISS [OIII]/H$\beta$ spectroscopy.
The right panel of Figure~\ref{Spec} shows the extracted 1D/2D grism spectra of \targetname. The grism spectra do not show obvious emission lines, while the 4,000\,\AA\ break is tentatively detected in F150W band. The best-fit redshift based on the grism spectra provided by \textsc{Grizli} is consistent with that determined by ALMA (see the probability density distribution in Figure~\ref{Spec}).

\subsection{Size Ratio between Dust and Stellar Components}
\label{sub_sec_res}

Dust in the central few kpc region is likely generated by a central starburst, favoring an inside-out star-forming scenario \citep{Nelson2016, Tadaki2020}.
Meanwhile, the rest-frame optical continua allows to probe the galaxy-scale star formation \citep{Gullberg2019}.
Thus, the dust-to-stellar effective radius ratio ($R_{\rm e, dust}$/$R_{\rm e, star}$) reflects the comparison between two-component star-formation models \citep{Lang2019, Tadaki2020, Sun2021}.
This dusty spiral galaxy, \targetname, provides us the opportunity to connect the intense star-formation process and the structure formation of spiral arms and bulges.

To investigate the stellar component, we used reconstructed broad-band imaging observations that trace the rest-frame optical emission of \targetname.
However, for star-forming galaxies, the size measured from the stellar density map would be more compact than that observed from surface brightness, which is due to the surface brightness dimming effects \citep{Wuyts2012, Genzel2020}.
Therefore, measuring half-light radius would overestimate the size of stellar mass. 
Fortunately, there is a strong empirical relation between stellar mass-to-light ($M_{\rm star}/L$) ratio and the optical colors for spiral galaxies \citep[e.g.,][]{Bell2001}.

\begin{figure*}
    \centering
	\includegraphics[width=0.95\textwidth]{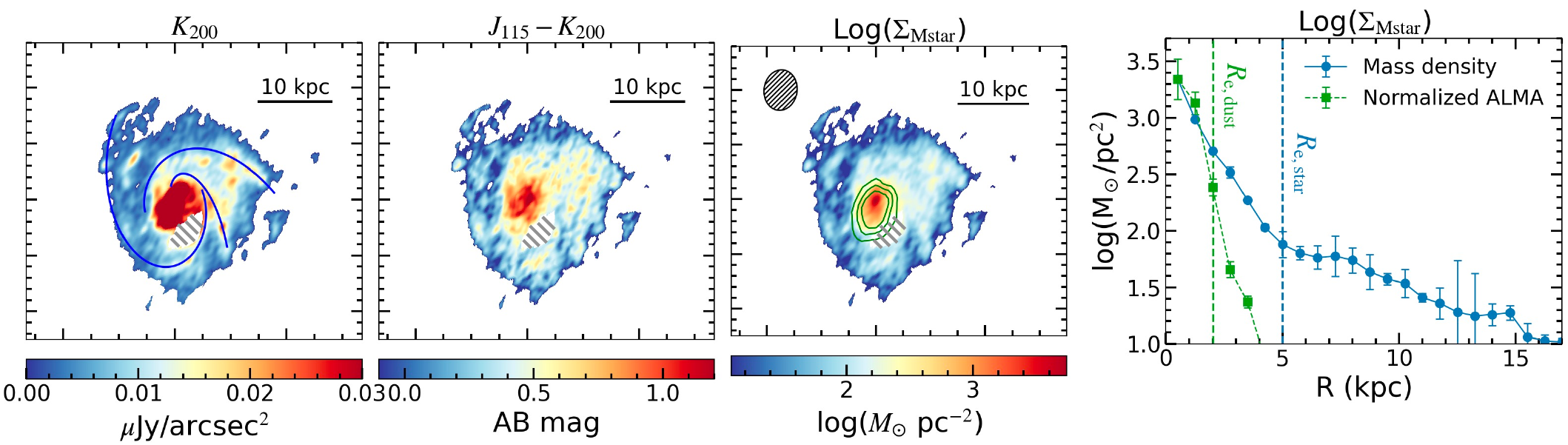}
	\caption{ 
	{\bf From left to right}: Reconstructed $K_{\rm 200}$-band cutout map, $J_{\rm 115} - K_{\rm 200}$ color map, stellar surface density map, and lensing-corrected  stellar-density/dust-continuum radial profile. Pixels belonging to \targetname\ with fluxes higher than $2\sigma$ (determined by the pixel-pixel standard deviation) were shown. The stellar mass surface density map is derived based on a best-fit $M_{\rm star}/L$ vs. color relation. In the $K_{\rm 200}$-band map (Left panel), blue lines represent spiral arms fitted by the Logarithmic spirals equation (see Section~\ref{sub_sec_res}, Eq.~\ref{spiral_arm}). Additionally, the ALMA dust-continuum observations are shown in the stellar surface density map (the third panel) as green contours ranging from [5, 7.5, 10]$\times \sigma$. In the right panel, we obtained the effective radius of stellar mass and dust continuum from the observed radial profile, i.e., $R_{\rm e, star}=5.0\pm1.5$ kpc, $R_{\rm e, dust}=2.04 \pm 0.26$ kpc. Tentative foreground interloper is masked as grey lines.}
	\label{stellar_mass_map}
\end{figure*}

Following \citet{Lang2019}, we derived the stellar mass based on broad-band color ($J_{\rm 115}-K_{\rm 200}$) .
Our photometric process is listed as follows.
Firstly, we matched point-spread functions (PSFs).
To construct PSFs in different filters, we selected stars from Gaia DR3 archival catalog \citep{Gaia2016, Gaia2022} and then stacked unsaturated ones in the field-of-view. 
After matching PSFs to that of F200W,
we performed pixel-pixel photometry in both $J_{\rm 115}$ and $K_{\rm 200}$ bands.
The photometric zeropoints are determined using the calibration file
``\texttt{jwst\_niriss\_photom\_0028.rmap}''.
The $K$-band observations and color maps are shown in the left two panels of Figure~\ref{stellar_mass_map}.
In Figure~\ref{stellar_mass_map}, the spiral associated pixels were identified with the flux density over $2\sigma$ noise (determined by the pixel-to-pixel standard deviation). 
{Here, we note that there could be a foreground target that could contaminate \targetname\ (See Appendix A).}
Thus, we masked it before performing measurements. 

We then obtained the stellar mass density ($\Sigma_{\rm M_{star}}$) map based on the relation between mass-to-light ratio and color.
This relation was constructed and calibrated using simulated spectral energy distributions (SEDs) derived from stellar population synthesis (SPS) methods (see Appendix B).
The best-fit relation is:
\begin{equation}
    M_{\rm star}/L = 0.55\times(J_{\rm 115} - K_{\rm 200}) + 18.92,
\end{equation}
where the mass and luminosity are in solar units. We then obtained the pixelated $\Sigma_{\rm M_{star}}$ map (third panel of Figure~\ref{stellar_mass_map}) with the $M_{\rm star}/L$--color relation above.
{
We measured the radial profile of stellar surface density by averaging it in the deprojected galactocentric plane with the radial interval of 1\,kpc (see the last panel of Figure~\ref{stellar_mass_map}).}
The major axis of the galaxy was determined by the second-order moments of its stellar mass distribution relative to its mass density peak \citep{Chen2021}.
{
The $R_{\rm e, star}$, defined as the radius containing half mass, was calculated from the $\Sigma_{\rm M_{star}}$ profile.}
The measured $R_{\rm e, star}$ is $5.0\pm1.5$ kpc.
We note that the $R_{\rm e, star}$ was calculated by numerically integrating the mass density profile out to 17 kpc, approaching the outskirt of the disk.

The effective radius of the dust continuum at a rest-frame wavelength of 280\,\micron\ was measured using ALMA Band\,6 data (2018.1.00035.L, PI: Kohno) through a UV-plane visibility profile modeling. 
The best-fit circularized effective radius assuming a Gaussian profile is $R_\mathrm{e,dust}=0.42\pm0.05\arcsec$, corresponding to a physical size of 2.04 $\pm$ 0.26 kpc after correcting for lensing magnification.
This is consistent with the size derived assuming an exponential profile (0.39 $\pm$ 0.07$\arcsec$), and the detailed methodology will be presented by Sun et al.\ (in prep.) among a larger sample of lensed dusty star-forming galaxies discovered with the ALCS \citep{Sun2022}.
We conclude here that the measured radius ratio of \targetname\ is $R_{\rm e,dust}/R_{\rm e,star} = 0.41\pm0.13$. 
{We also note that detailed CO kinematic information is provided in the Appendix C.}

Furthermore, we measure the physical parameters of the observed spiral-arm like features.
Following \citet{Law2012}, we used the Logarithmic spiral equation \citep{Davis2012}: 
\begin{equation}
    r=r_{0}e^{\theta \ {\rm tan (\phi)}},
    \label{spiral_arm}
\end{equation}
where $r,\ r_{0},\ \theta$, and $ \phi$ represent the radius, initial radius, angle, and pitch angle of arms, respectively.  
In the NIRISS-F200W imaging data, there are three spiral arms flowing into the red bulge. 
Thus, considering inclination effects, we fit their shape using Equation~\ref{spiral_arm}.  
The best-fit results (blue lines in the left panel of Figure~\ref{stellar_mass_map}) show that the logarithmic equation effectively reproduces the arms. 
The best-fit pitch angle of the arms is $\phi=34\pm13$\arcdeg, similar to that of the three-armed spiral galaxy at $z=2.18$ \citep{Law2012}. 

\section{discussion} \label{sec:discussion}
\subsection{Implication of the Observed Dust-to-stellar-radius Ratio}

{Smaller dust-to-stellar-radius ratio ($R_{\rm e,dust}/R_{\rm e,star}$) indicates more compact dust emission which is possibly caused by gas-rich merger and more active star formation \citep[]{Tacchella2016, Nelson2016, Gullberg2019}. In contrast, larger $R_{\rm e,dust}/R_{\rm e,star}$ ratio corresponds to relatively weak star formation (\citealt{Tadaki2020,Sun2021}).}
To analyze the star-forming and structural
properties, following \citet{Tadaki2020} and \citet{Sun2021},
{we compare the main sequence offset ($\Delta$MS) and the dust-to-stellar size ratio with those of dusty star-forming galaxies at $z\sim 2 $. }
$\Delta$MS is defined as the star-formation rate (SFR) ratio between the observed value and that expected on the so-called star-forming ``main sequence" (SFR$_{\rm MS}$), i.e., $\Delta$MS $=\log$[SFR/SFR$_{\rm MS}$] \citep{Speagle2014}. 
The observed SFR of \targetname\ is $\log$[SFR$_{\rm obs}/(\mu^{-1}M_{\odot}\ {\rm yr}^{-1})]=2.32 \pm 0.16$ derived by SED-fitting results and reported in \citet{Sun2022}.
We follow \citet{Speagle2014} to calculate the SFR$_{\rm MS}$ (SFR$_{\rm MS}$ = SFR$_{\rm MS}(M_{\rm star},z)$), using the stellar mass of \targetname\ by summing up the de-lensed stellar-mass density map, which is $\log(M_{\rm star}/M_{\odot})=10.55$.
With an estimated main-sequence SFR of $\log[$SFR$_{\rm MS}/(M_{\sun} {\rm \ yr^{-1}})]$ of $\approx2.01$ (de-lensed), the intrinsic $\Delta$MS of \targetname\ is $\approx -0.08$. 

\begin{figure}
    \centering
	\includegraphics[width=0.48\textwidth]{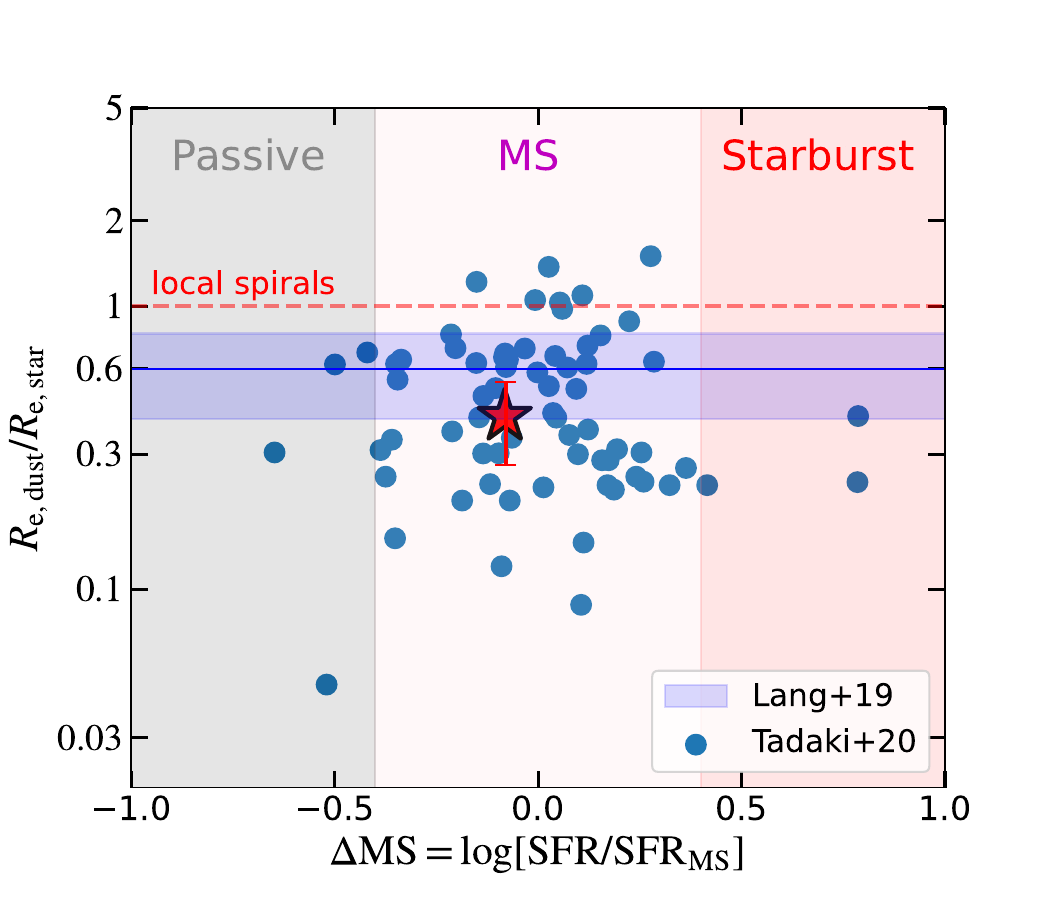}
	\caption{Dust and stellar effective radius ratio vs. main sequence offset ($\Delta$MS). Blue dots are massive SFGs at $z\sim2$ reported by \citet{Tadaki2020}, while the red star represents the results of \targetname. Shaded blue region indicates the radius ratio observed of SMGs at $z\sim2$ \citep{Lang2019}. $\Delta$MS is defined as SFR ratios between the observed value and the expected ones on the star-forming MS. Following \citet{Aravena2020}, the boundaries between passive, main sequence, and starburst galaxies are distinguished by $\Delta $MS $=\pm 0.4$, respectively. We conclude that, the spiral galaxy, \targetname, having a more compact dust core compared with SMGs at $z\sim2$ and local spirals. And the star-forming intensity is the same as main-sequence galaxies at $z=3$. 
	{Our observations suggest that for \targetname\ at $z=3$, the bulge was formed with  star formation at the center core region ($\lesssim2$ kpc). }}
	\label{R_ratio_M}
\end{figure}

{We compare our size measurements with those of dusty star-forming galaxies at $z\sim2$ \citep{Lang2019, Tadaki2020} in Figure~\ref{R_ratio_M}.}
{The average $R_{\rm e, dust}/R_{\rm e,star}$ of SMGs is $0.6\pm 0.2$ (blue shaded region in Figure~\ref{R_ratio_M}; \citealt{Lang2019}), suggesting that the distribution of dust components is more compact than the stellar ones \citep[see also][]{Hodge16,Sun2021,GG22}.}
This result is also consistent with that observed from massive star-forming galaxies (blue dots in Figure~\ref{R_ratio_M}; \citealt{Tadaki2020}).
As an SMG, the observed radius ratio (0.41) of \targetname\ is {broadly consistent with SMGs in previous studies, but smaller than} those of local spirals ($R_{\rm e, dust}/R_{\rm e,star}\approx 1$; \citealt{Hunt2015,Bolatto17}).
{Given the observed $\Delta$MS ($-0.08$), our results suggest that, as a spiral galaxy, \targetname\ is on the star-forming main sequence ($-0.4<\Delta{\rm MS} < 0.4$), similar to local spirals that are also on the MS. However, \targetname\ hosts a more compact dust core than those of local spirals while consistent with the majority of SMGs at $z\approx2$ (Figure~\ref{R_ratio_M}). }

{The majority of observed SFR of \targetname\ at the current epoch is obscured and traced by the compact FIR emission at the center core region ($R_{\rm e,dust}\approx 2$ kpc), further surrounded by a more extended stellar disk ($\approx 5$ kpc).}
{The observed dusty core size ($\approx2$ kpc) of A2744-DSG-z3 is comparable to the typical galactic bulge size ($\approx 1$ kpc) for galaxies with stellar mass of $\sim10^{11} M_{\odot}$ at $z\sim2$ \citep{Tadaki2017}. 
The observed star formation at the bulge scale suggests an active bulge formation scenario.}
{
After the quenching of the compact dusty star formation in the core of \targetname, a compact stellar component will remain in the center of the galaxy as the newly formed bulge.
}

\subsection{Group Environment and the Formation of Spiral-like Structure at $z\gtrsim3$}

Thanks to the JWST/NIRISS grism observations, the target close to \targetname\ ($GLASS-Z$grad1, recently reported in \citealt{Wang2022}) is identified as one companion galaxy at the same redshift.
The stellar mass of $GLASS-Z$grad1 is $\approx 10^9\ M_{\odot}$, and the de-lensed angular separation between the two soruces is $\approx 2\arcsec$, corresponding to 15.4\,kpc at $z=3.06$.
The stellar mass of \targetname\ is $\approx 10^{10.6} M_{\odot}$, suggesting a halo mass of $\sim 10^{12.5}\ M_{\odot}$ \citep{Rodriguez_Puebla2017,Somerville2018} with a Virial radius of $\sim 120$ kpc 
\footnote{\href{https://github.com/ylu2010/DarkMatterHaloCalculator}{https://github.com/ylu2010/DarkMatterHaloCalculator}}.
The lensing corrected projected separation between these two galaxies reside within the Virial radius. 
With the mass ratio between these two galaxies being $\approx 36$, this implies a minor-merger interaction scenario that potentially triggered the dusty central starburst in \targetname.
\citet{Law2012} also suggest that one spiral galaxy at $z\sim2$ could be formed by the minor merger process.

We note that, in the field A2744, \citet{Boyett2022} reported a number counts excess at $z\sim3$. 
Thus, \targetname\ could reside in an overdense region. 
For the companion galaxy, $GLASS-Z$grad1, the measured slope in the metallicity distribution is $\Delta \log({\rm O/H})/\Delta r = 0.165\pm 0.022$ \citep{Wang2022}. 
The occurrence of inverted gradients measured in
this galaxy could also be explained by an ongoing minor merger. Due to the loss of angular momentum by the torque in the merger interactions, the metal-poor gas inflows into the inner disk to flatten/invert the metallicity gradient \citep{Krabbe2011}. Alternatively, the group environment may also give rise to enhanced cold mode accretion on the member galaxies, which inverts the metal gradient \citep{Li2022}.

To understand the environmental influence on the spiral formation, firstly, we note that the stellar spiral in \targetname\ is present at $z = 3.06$, just around the peak of cosmic star formation (see \citealt{MD14} for a review). 
Thus, the number of galaxies and merger rates could be high.
{Cosmological simulations suggest that high-redshift spiral structures may be caused by higher merger rate} \citep{Hammer2009}, high inflow rate, and the low angular momentum of cold streams \citep{Cen2014}. 
Meanwhile, there are simulations reporting that the spirals exist by $z\approx 3$ \citep{Fiacconi2015} and the high-redshift spirals may be arisen from swing amplifications triggered by galaxy interactions.
\citet{Kohandel2019} further showed that the cold gaseous spiral structure (close to the reionization epoch) can exist when the galactic disk is relaxed after a merger. 
{\targetname\ could experience a merger-like process;} and this could enhance the formation of spirals as predicted by cosmological simulations.

\section{Summary} \label{sec:summary}

In this Letter, we present the first identification of a dusty spiral galaxy (\targetname) at $z=3.059$ observed by ALMA and JWST/NIRISS in both slitless-spectroscopic and direct imaging modes. 
The redshift of \targetname\ is determined through ALMA CO observations. 
Meanwhile, by fitting the rest-frame optical continuum spectra provided by NIRISS/girsm using \textsc{Grizli}, we also obtained a redshift solution of \targetname\ with $z\approx 3.07$, consistent with that obtained from ALMA observations.
The best-fit three spiral-arm models to the NIRISS/F200W direct images suggests an inclination/pitch angle of $28$ (deg) and $34\pm13$ (deg), respectively. 
We convert the observed F115W--F200W colors and flux densities to stellar masses, and find a stellar-mass effective radius of $5.0$ kpc, while the effective dust radius is $2.04\pm0.26$ kpc.
Furthermore, the NIRISS grism data also identify a companion dwarf galaxy, called $GLASS-Z$grad1 \citep{Wang2022}, with the projected separation of 15.4 kpc (de-lensed). 
The identification of this galaxy pair suggests a minor-merger scenario that triggers the dusty central starburst in \targetname.

The high spatial-resolution imaging observations provided by JWST give us the first opportunity to resolve the detailed structure of a spiral galaxy at $z\sim3$ to a physical scale of $\approx290$ pc.  
Additionally, the NIRISS grism spectroscopic observations provides us the chance to determine redshifts for bright targets or emitters within a sky coverage of $\sim 2\arcmin$. 
Like the galaxy pair identified in this work, spectroscopic redshifts through emission lines will further help us to investigate the environments of galaxies at different redshifts. 

The observed dust vs. stellar component radius ratio (0.41) of \targetname\ suggests that the dust core is more compact than that observed of local spirals ($\approx1$). 
The observed main sequence offset (-0.08) indicates that this target is undergoing the same intense star-formation activities as main-sequence galaxies at $z=3$. 
Furthermore, 
{
A possible minor-merger scenario may help to explain
}the observed inverted metallicity gradient of $GLASS-Z$grad1 due to the interaction. 
We note that, future high spatial-resolution observations (e.g, ALMA CO for \targetname\ and JWST/NIRSpec-IFU for $GLASS-Z$grad1) 
will allow to probe the dynamics of this minor-merger candidate.

\section*{Acknowledgments}

Y.W. thanks Shude Mao, Jerry Sellwood, Zuyi Chen, Xin Wang and for very helpful discussions. 
We thank the anonymous referee for reading the paper carefully and providing comments that helped improve and strengthen this paper.
We thank the data editor, for carefully helping us check the used data and software.
Z.C., Y.W., X.L., Z.L., M.L., J.L., \& S.Z. are supported by the National Key R\&D Program of China (grant no.\ 2018YFA0404503) and the National Science Foundation of China (grant no.\ 12073014).
F.S.\ acknowledges support from the NRAO Student Observing Support (SOS) award SOSPA7-022. 
F.S.\ and E.E.\ acknowledge funding from JWST/NIRCam contract to the University of Arizona, NAS5-02105. 
FEB acknowledges support from ANID-Chile BASAL CATA ACE210002 and FB210003, FONDECYT Regular 1200495 and 1190818,
and Millennium Science Initiative Program  – ICN12\_009.
F.W.\ is thankful for support provided by NASA through the NASA Hubble
Fellowship grant no. HST-HF2-51448.001-A awarded by the Space Telescope Science Institute, which is operated by the Association of Universities for Research in Astronomy, Inc., under NASA contract NAS5-26555. 

This work is based on observations made with the NASA/ESA/CSA James Webb Space Telescope. The data were obtained from the Mikulski Archive for Space Telescopes at the Space Telescope Science Institute, which is operated by the Association of Universities for Research in Astronomy, Inc., under NASA contract NAS 5-03127 for JWST. These observations are associated with program ERS-1324.
The authors acknowledge the GLASS team for developing their observing program with a zero-exclusive-access period.

This paper makes use of the following ALMA data: ADS/JAO.ALMA\#2017.1.01219.S. ALMA is a partnership of ESO (representing its member states), NSF (USA) and NINS (Japan), together with NRC (Canada), MOST and ASIAA (Taiwan), and KASI (Republic of Korea), in cooperation with the Republic of Chile. The Joint ALMA Observatory is operated by ESO, AUI/NRAO and NAOJ. In addition, publications from NA authors must include the standard NRAO acknowledgement: The National Radio Astronomy Observatory is a facility of the National Science Foundation operated under cooperative agreement by Associated Universities, Inc.

This work is based on data and catalog products from HFF-DeepSpace, funded by the National Science Foundation and Space Telescope Science Institute (operated by the Association of Universities for Research in Astronomy, Inc., under NASA contract NAS5-26555).

JWST data used in this paper were obtained from the Mikulski Archive for Space Telescopes (MAST) at the Space Telescope Science Institute. The specific observations analyzed can be accessed via \dataset[10.17909/91zv-yg35]{https://doi.org/10.17909/91zv-yg35}.

\vspace{5mm}
\facilities{JWST, ALMA, HST}

\software{astropy \citep{Astropy_Collaboration2022},  
          {\tt BAGPIPES} \citep{Carnall2018},
          {\tt glafic} \citep{Oguri2010},
          {\tt EAZY} \citep{Brammer2008},
          {\tt Qubefit} \citep{Neeleman2020}
          {\tt Grizli} \citep{Brammer_grizli_2022}}

\appendix

\section*{Appendix A: Tentative Foreground contamination} \label{Sec:Contamination}
\begin{figure*}
    \centering
	\includegraphics[width=0.9\textwidth]{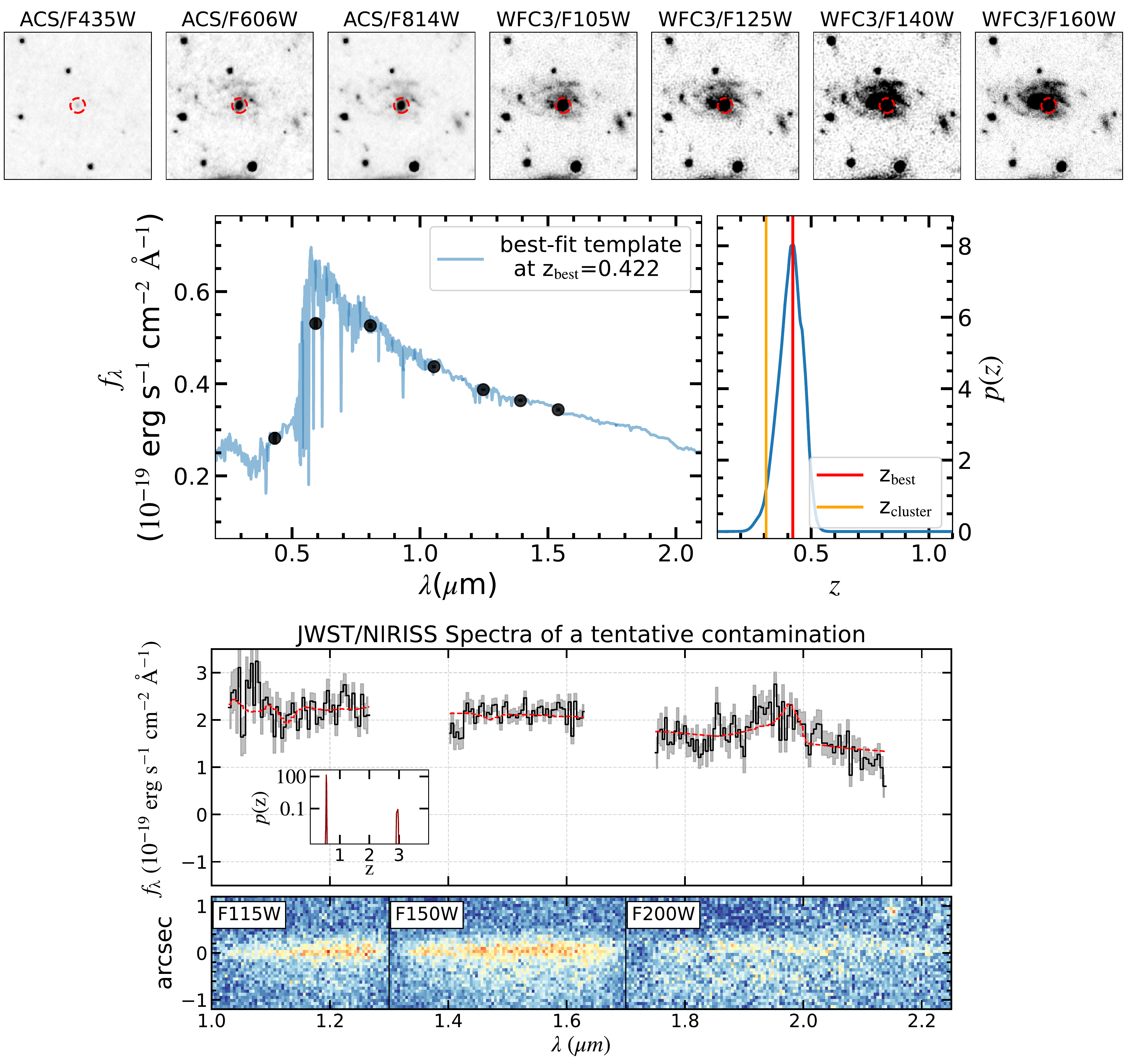}
	\caption{
    {\bf Top}: Seven-bands HST observations. The red cycle indicates the location and apertures ($r=0.5\arcsec$) used for this tentative foreground galaxy.
    {\bf Middle}: Photometric redshift solution based on \textsc{EAZY}. In the left panel, the best-fit SED is shown in blue. Meanwhile, in the right panel, the redshift probability distribution is shown in blue. The best-fit redshift ($z_{\rm best}=0.422$) and the redshift of cluster A2744 are marked as red and orange lines, respectively.
    {\bf Bottom}: 2D/1D JWST/NIRISS spectra of the tentative target. Each panel is the same as that in the right panel of Figure~\ref{Spec}. We conclude that, from the grism spectra, we cannot directly determine the redshift of the tentative contamination because there are no obvious emission lines. The best-fit redshift obtain from \textsc{Grizli} is $z=0.544^{+0.006}_{-0.003}$. However, there is still a high-redshift solution. Thus, we cannot rule out the possibility that this target is located at the same redshift of \targetname.}
	\label{contamination}
\end{figure*}

In Figure~\ref{spiral_arm_rgb}, there is a white core near the center of \targetname. 
In this section, we discuss the possibility of contamination caused by a foreground cluster dwarf galaxy. 
We combined archival HST photometric observations and JWST/NIRISS grism data to reveal whether it is a foreground {galaxy}. 
Figure~\ref{contamination} shows these observations. 
We performed {photometric measurements} using an aperture with radius of $0.5\arcsec$ on the PSF-matched images obtained from the HFF-DeepSpace \citep{Shipley2018,Nedkova2021}.
Based on SED-fitting results, the best-fit redshift of this target is $z=0.422$ ($z=0.544^{+0.006}_{-0.003}$) given by \textsc{EAZY} (\textsc{Grizli}).
We conclude that from the photometric and grism data, this target could be a foreground contamination, {despite that the redshift determination of this target is not secure}. 

\section*{Appendix B: Mass-to-light Ratio vs. Color Relation} 
Empirically, \citet{Bell2001} presented a strong correlation between stellar mass-to-light ($M_{\rm star}/L$) ratio and the optical colors for spiral galaxies.
Therefore, the JWST rest-frame optical and high spatial resolution observations give us the first opportunity to resolve the stellar mass distribution of a spiral galaxy at $z=3.06$. 
To measure the pixelated stellar mass of \targetname, we follow the procedures in \citet{Lang2019}.
The $M_{\rm star}/L$ vs.\ color relation is constructed and calibrated by synthetic galaxy SEDs. 

\begin{figure*}
    \centering
	\includegraphics[width=0.48\textwidth]{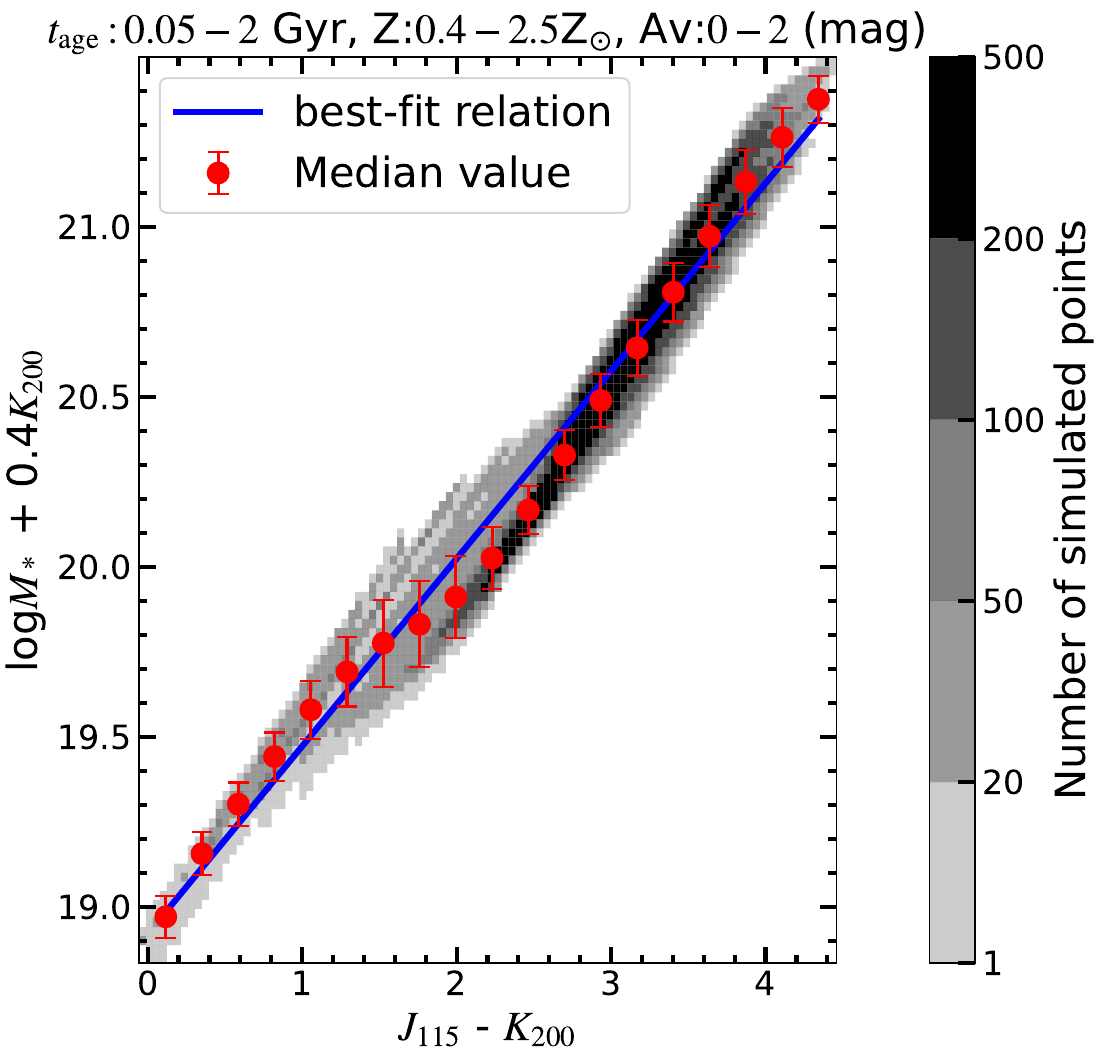}
	\caption{Simulated $M_{\rm star}/L$ vs.\ color relation. Gray-scale images represents the number of simulated data points with different stellar parameters (e.g., $t_{\rm age}$, Z, and ${\rm A_v}$). Stellar ages range from 50 Myr to 2 Gyr at $z=3.06$, while metallicity ranges from 0.4 to 2.5 Z$_{\odot}$. The simulated colors are derived by adopting the \citet{Calzetti2000} law with ${\rm A_V} = 0.0 - 2.0$ (mag). Red points are median values in different color bins, while error bars are standard deviations in each bin. The blue line is a best-fit linear relation using the median points.}
	\label{C_MtL}
\end{figure*}

To generate mocked SEDs, we use a public python package {\tt BAGPIPES} \citep{Carnall2018} by adopting different stellar parameters (e.g., the age/metallicity of stellar populations $t_{\rm age}$/$Z$, and the dust attenuation parameter in the rest-frame V band, $A_{\rm V}$).
For galaxy models, we assume a \citet{Kroupa2002} initial mass function with $t_{\rm age}$ ranging from $0.05-2$ Gyr, while $Z$ ranges from $0.4-2.5$ $Z_{\odot}$. Then, we assume an exponentially declining star formation history (SFR) with timescales of $\tau_{\rm SFR}  = 30 $ Myr, which is guided by the best fitting result of a similar red spiral galaxy at $z\sim2.5$ \citep{Fudamoto2022}. 
In order to convert these mocked galaxy SEDs to colors, we apply a \citet{Calzetti2000} dust-attenuation law with $A_{\rm V} = 0.0 - 2.0$.
Fig.~\ref{C_MtL} shows the $M_{\rm star}/L$ vs. color relation. We define the $M_{\rm star}/L$ as ($\log M_{\rm star} + 0.4K_{\rm 200}$), similar to that in \citet{Lang2019}. The best-fit linear relation is $M_{\rm star}/L = 0.55\times(J_{\rm 115} - K_{\rm 200}) + 18.92$.

\section*{Appendix C: CO moment maps}

\begin{figure*}
    \centering
	\includegraphics[width=0.9\textwidth]{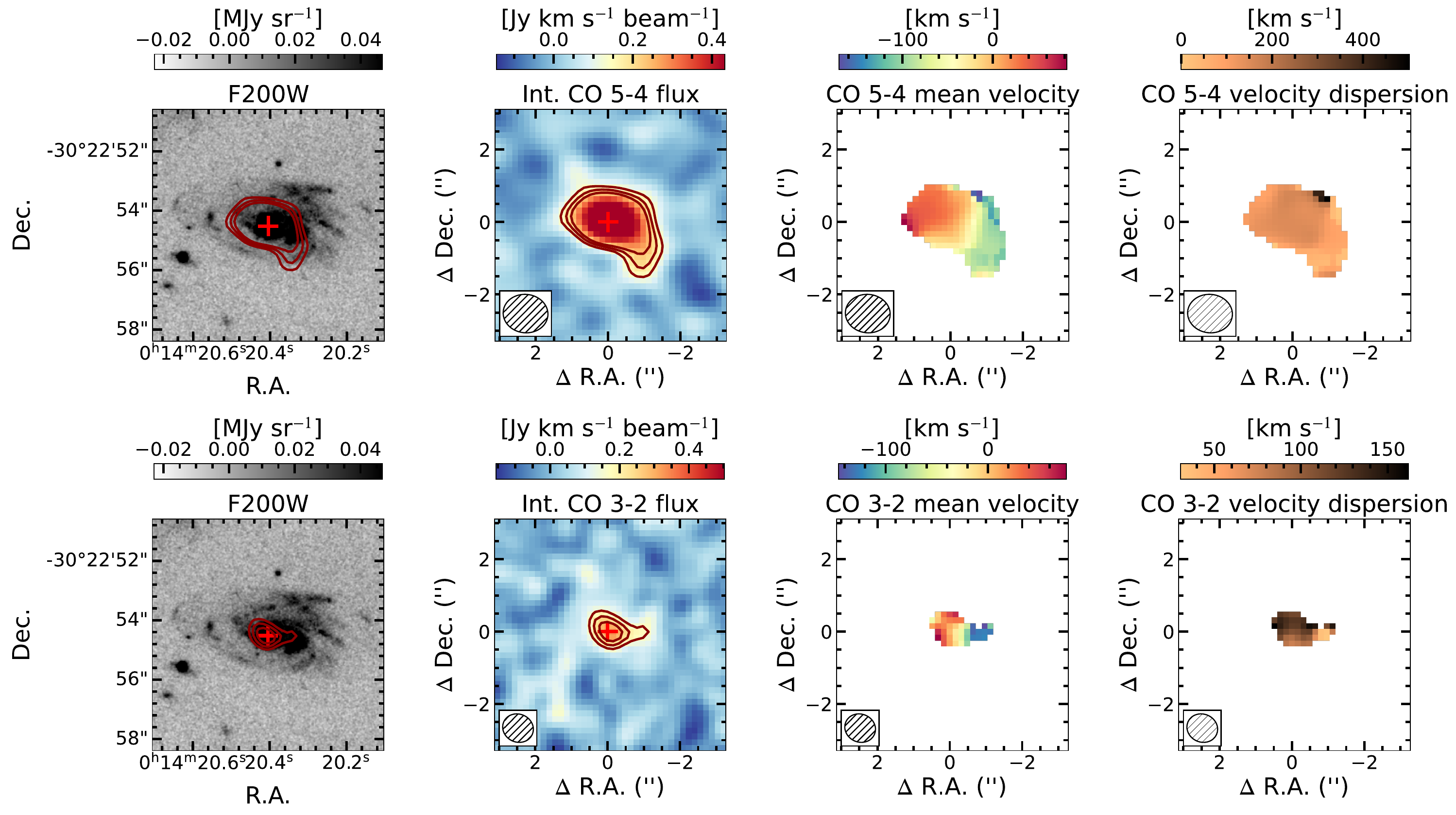}
	\caption{{\bf Left}: JWST/NIRISS F200W observations, overlaid with contours of the CO lines (in darkred). The red cross represents the center of \targetname. 
	{\bf Second to last column}: CO flux density, velocity and velocity dispersion. The synthesized beams are shown at the lower left corner of each panels with size of $1.24\arcsec \times 1.06\arcsec$ (CO (5-4)) and $0.86\arcsec \times 0.77\arcsec$ (CO (3-2)), respectively. The contours are drawn at [3, 4, 5]$\times$ 1$\sigma$. For mom-1 and mom-2 maps, regions with flux smaller than $2\sigma$ are masked.
	}
	\label{CO_mom_map}
\end{figure*}

We extract the {kinematic information} from the CO data cubes by constructing velocity maps (mom-1) and velocity dispersion maps (mom-2).
Here, we {display} the results in Figure~\ref{CO_mom_map}.
For \targetname, using the CASA task {\tt imfit}, the CO intensity is marginally resolved with the de-convolved sizes of
$1.43\arcsec \times 0.66\arcsec$ and $1.37\arcsec \times 0.79\arcsec$ for CO (5-4) and CO (3-2), respectively.
This corresponds to a physical size of $4.5{\rm\ kpc}\ \times 2.1{\rm\ kpc}$ and $ 4.3 {\rm\ kpc} \times 2.5 {\rm\ kpc}$ at $z=3.059$ (de-lensed).
The measured size is slightly smaller than those of five star-forming galaxies with comparable mass at similar redshifts \citep{Cassata2020}.
Moreover, the CO mom-1 maps show a tentatively rotating feature (the third-column two panels in Figure~\ref{CO_mom_map}). 
The measured rotation velocity is about $\sim200$ km s$^{-1}$ (inclination corrected).

\bibliography{sample631}{}
\bibliographystyle{aasjournal}

\end{document}